\documentclass[12pt]{article}
\usepackage{amsfonts}
\usepackage{psfig}

\begin{document}

\newcommand{\R}{{\mathbb R}}
\newcommand{\C}{{\mathbb C}}
\newcommand{\Z}{{\mathbb Z}}
\newcommand{\be}{\begin{equation}}
\newcommand{\ee}{\end{equation}}
\newcommand{\bea}{\begin{eqnarray}}
\newcommand{\eea}{\end{eqnarray}}
\newcommand{\tb}{\tilde{\beta}}
\newcommand{\ta}{\tilde{a}}
\newcommand{\tal}{\tilde{\alpha}}
\newcommand{\ttau}{\tilde{\tau}}
\newcommand{\td}{\tilde{\delta}}
\newcommand{\cQ}{{\mathcal{Q}}}
\newcommand{\tQ}{{\tilde{\cQ}}}
\newcommand{\cF}{{\mathcal{F}}}
\newcommand{\cG}{{\mathcal{G}}}
\newcommand{\tG}{{\tilde{\cG}}}
\newcommand{\tN}{{\tilde{N}}}
\newcommand{\tF}{{\tilde{F}}}
\newcommand{\tV}{{\tilde{V}}} 
\newcommand{\cH}{{\mathcal{H}}}
\newcommand{\cX}{{\mathcal{X}}}  
\newcommand{\cY}{{\mathcal{Y}}}
\newcommand{\cR}{{\mathcal{R}}}
\newcommand{\cT}{{\mathcal{T}}}
\newcommand{\cA}{{\mathcal{A}}}  
\newcommand{\lt}{\frac{\Lambda}{3}}
\newcommand{\cM}{{\mathcal{M}}}

\newcommand{\nn}{\nonumber}
\newcommand{\tA}{\tilde{A}}
\newcommand{\tp}{\tilde{\phi}}
\newcommand{\e}{\varepsilon}
\newcommand{\mbin}{\mbox{in}}
\newcommand{\mbout}{\mbox{out}}

\newcommand{\az}{\alpha}
\newcommand{\bz}{\beta}
\newcommand{\cz}{\gamma}
\newcommand{\gz}{\gamma}
\newcommand{\dz}{\delta}
\newcommand{\ez}{\epsilon}
\newcommand{\kz}{\kappa}
\newcommand{\lz}{\lambda}
\newcommand{\na}{\nabla}

\newcommand{\ou}{\tilde{u}}
\newcommand{\on}{\tilde{n}}
\newcommand{\oN}{\tilde{N}}
\newcommand{\oV}{\tilde{V}}
\newcommand{\ok}{\tilde{k}}
\newcommand{\oa}{\tilde{a}}
\newcommand{\oep}{\tilde{\varepsilon}}
\newcommand{\oj}{\tilde{\jmath}}
\newcommand{\os}{\tilde{s}}

\newcommand{\bT}{\underline{T}}
\newcommand{\bI}{\underline{I}}

\baselineskip0.6cm

\pagestyle{plain}
\pagenumbering{arabic}

\title{Moving Observers, Non-orthogonal Boundaries, and Quasilocal Energy}
\author{I.S. Booth\footnote{ivan@avatar.uwaterloo.ca} and R.B. 
Mann\footnote{mann@avatar.uwaterloo.ca} \\
        Department of Physics\\
        University of Waterloo\\
        Waterloo, Ontario\\
        N2L 3G1}
\date{October 1, 1998\\}
\maketitle
\begin{abstract}
The popular Hamilton-Jacobi method first proposed by Brown and York for
defining quasilocal quantities such as energy for spatially bound regions
assumes that the timelike boundary is 
orthogonal to the foliation of the spacetime. Such a restriction is
undesirable for both theoretical and computational reasons. 
We remove the orthogonality assumption and show that it is more natural
to focus on the foliation of the timelike boundary rather than the 
foliation of the entire four dimensional bound region. Reference spacetimes which
define additional terms in the action are discussed in detail. 
To demonstrate this new formulation, we calculate the quasilocal energies
seen by observers
who are moving with respect to a Schwarzschild black hole. 
\end{abstract}

\section{Introduction}

Gravitational thermodynamics and its relationship to the
Euclidean-action formulation of quantum
gravity have been of increasing interest in recent years.  
This relationship was first explored by 
Gibbons and Hawking \cite{GibH}, who argued that the Euclidean
gravitational action is equal to the grand canonical free energy times
the reciprocal of the temperature associated with a black hole (or
cosmological) event horizon \cite{BekH}. A more recent
extension of this work by Brown and York involved consideration 
of the formulation of the
partition function for gravitating systems of finite spatial extent
\cite{BY1,BY2}. Starting from a spacelike foliation of a finite region of
spacetime, and a timelike vector field defining a flow of time, they 
studied the Einstein-Hilbert action 
using a Hamilton-Jacobi type analysis. Decomposing the action according to 
the foliation and flow of time, they showed that natural candidates arose 
for quantities such as energy  and momentum. These quantities were
defined quasilocally, i.e. for a region of finite spatial extent containing
a gravitating system.

For a number of reasons this analysis and its associated quasilocal
quantities have generated much interest and found a multitude of uses.
First, all physical systems with which we have any experience have a
finite spatial boundary. Indeed one of the central concepts in
thermodynamics is that of a system and a reservoir that are separated by a
partition. Quasilocal quantities admit a physical realization of these
concepts so that thermodynamics may be applied in a sensible way. As such,
in the literature this analysis has been used extensively in the study of
black hole thermodynamics (for example in \cite{BY2,bht,BCM}). 
Among other
places, this work has found application in studies of the
distribution of gravitational energy in a variety of spacetimes (for
example \cite{examples}) and also in examining the quantum mechanical
creation of pairs of black holes (for example \cite{pc}). A very similar
Hamiltonian decomposition of the action has also been executed by Hawking
and Horowitz \cite{HHorowitz}. 

However an acknowledged incompleteness exists in the quasilocal 
formulation in that (apart from two exceptions mentioned below) the
spacetime
foliation is always assumed to be orthogonal to the timelike boundary. While
this is the case for many standard examples (such as black holes surrounded
by a set of stationary observers) it is a fairly strong restriction. 
For example, within the confines of this orthogonality assumption it is
extremely difficult to calculate the quasilocal quantities seen by
observers who are falling into a black hole.
Furthermore when one considers variations of the metric 
(as one actually does during the quasilocal Hamilton-Jacobi analysis) 
the orthogonality assumption implies that the variations are not 
general, but instead restricted to those that 
preserve the orthogonality.

The requirement that the timelike boundary of the finite region be 
orthogonal to its spacelike boundary was dropped by Hayward \cite{Hayward}, who
considered how the basic Hilbert action $I$ should be modified so that
solving $\delta I = 0$ for  
general variations of the metric (subject to the boundary condition 
that boundary metrics 
should be held constant) will produce the Einstein equations in the 
usual way. However
this was from a purely Lagrangian viewpoint -- no consideration was given
as to how these variations would decompose in accordance with the 
spacetime foliation. 
That approach was recently considered by Hawking and Hunter 
\cite{HHunter} and Lau
\cite{Lau}, who addressed the 
non-orthogonal situation from a Hamiltonian perspective. 

In ref.\ \cite{HHunter} Hayward's action was broken down according to the 
foliation of the spacetime, a Hamiltonian proposed, and two sample 
calculations performed 
where the boundaries were non-orthogonal. However
there was no attempt made in this treatment
to consider the variation of the action and show an 
agreement between the quasilocal quantities suggested by that 
approach and the direct Hamiltonian deconstruction of the action. Furthermore,
in order to deal with the non-orthogonal intersections, the authors found it 
necessary to impose somewhat
complicated restrictions on background comparison spacetimes.

Lau's main interest in \cite{Lau} was to
reformulate the 
quasilocal quantities of \cite{BY1} in terms of Ashtekar 
variables. Treating non-orthogonal boundaries was a matter of secondary 
concern. Thus, although certain elements 
of his discussion are similar to some of the developments of this paper, 
the focus is quite different. In particular he did not decompose 
the action $I$,  
and the decomposition of
$\delta I$ was with respect to variations of the Ashtekar variables rather than 
metric variables. He did not discuss background terms in detail and did 
not calculate any examples. 

In this paper we shall consider both a decomposition of Hayward's action and 
a decomposition of the variation of that action and show that they 
agree in their natural candidates 
for the quasilocal quantities. In doing this, we shall focus on the boundary 
lapse and shift functions rather than the full spacetime lapse and shift 
as was 
the case in \cite{HHunter}. This will result in less complicated 
decompositions that also require less stringent restrictions on the comparison
spacetime, in contrast with the approach in ref.\cite{HHunter}. We shall also
argue that 
the boundary lapse and shift functions are the natural lapse and shift to consider. 

Before turning to those decompositions it is necessary to set out quite a
few definitions. Those definitions will be the subject of section two. In
section three we will perform the decompositions, examine the quasilocal
quantities that naturally arise from those decompositions, see how those
quantities relate to conserved charges, and finally examine the background
terms in some detail. Section four is made up of applications of the
theory of section three. In that section we shall calculate the quasilocal
energy (and other quantities) seen by a spherically symmetric 
set of observers undergoing a variety of motions in Schwarzschild spacetime.

\section{Definitions}

Consider a region $\cM$ of an $n$-dimensional spacetime with metric tensor
field $g_{\az \bz}$ and on that region define a timelike vector field
$T^\az$ and a spacelike $(n-1)$-dimensional hypersurface $\Sigma_0$. This
field and surface are sufficient to define a notion of time over $\cM$. As
a start, we let $\Sigma_0$ be an ``instant'' in time. That is we choose to
define all events happening on that surface as happening simultaneously. 
Next, consider a set of observers at locations $x^\az_{A0} \in \Sigma_0$
(where the $A$ index labels the individual observers). The past and future
locations of these observers are uniquely determined if we specify that
they must follow the paths $x^\az_A(t)$ through spacetime where these
paths satisfy the differential equation $\frac{d x^\alpha_A}{dt} = T^\az$
subject to the initial condition $x_A^\alpha(t_0) = x^\az_{A0}$. We then
define ``instants'' of time to be $t=\mbox{constant}$ surfaces. We label
them $\Sigma_t$ and define the notion of past and future by saying that if
$t_1 < t_2$, then $\Sigma_1$ ``happens'' before $\Sigma_2$. By
construction the $t_0$ hypersurface is $\Sigma_0$. Thus, from the vector
field and original hypersurface we have imposed an observer dependent
notion of time on our manifold according to the constructed time
coordinate $t$. Next, we may break up $T^\az$ into its components
perpendicular and parallel to the $\Sigma_t$ by defining a lapse function
$N$ and a shift vector field $V^\az$ so that,
\be
T^\az = N u^\az + V^\az,
\ee
where $u^\az$ is defined so that at each point in $\cM$ it is the 
future pointing unit normal vector to the appropriate 
hypersurface $\Sigma_t$, and $V^\az u_\az = 0$. The lapse and shift 
then tell us how observers being swept along with the time flow $T^\az$ 
move through space and time relative to the foliation. 

\begin{figure} 
\centerline{\psfig{figure=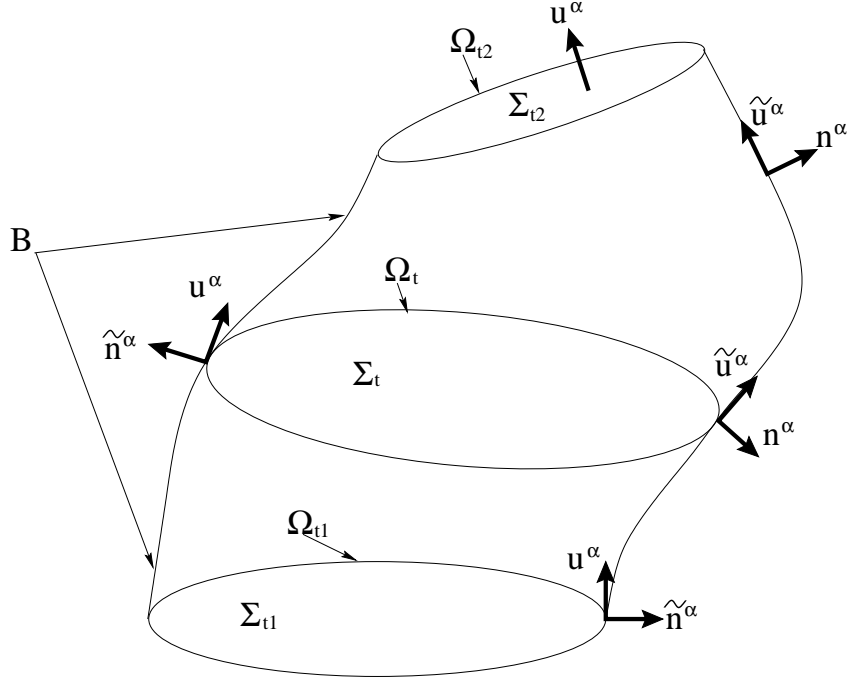,height=9cm,angle=270}} 
\caption{The
Lorentzian region $M$, its assorted normal vector fields, and a typical
element of the foliation.} 
\label{M} 
\end{figure}

We specialize to the situation of interest for this paper. On the surface
$\Sigma_0$ define a surface $\Omega_0$ that is topologically an $(n-2)$
sphere. $\Omega_0$ locally bifurcates $\Sigma_0$ - we will
pick one of the regions as ``inside'' and the other as ``outside''
\footnote{ Globally of
course the ``inside'' and ``outside'' could be connected. Consider for
example the case where $\Sigma_0$ is a two torus, and $\Omega_0$ is 
a homotopically non-trivial circle in that surface.}.  
Now, propagate this surface through time according
to the time flow $T^\az$. Then, by the local continuity of the time flow,
$\Omega_t$ will (at least locally) topologically remain an $(n-2)$ sphere
in the hypersurface
$\Sigma_t$ and still divide the ``inside'' from the ``outside''. Choosing
time coordinates $t_1$ and $t_2$ we define an $(n-1)$ dimensional
timelike hypersurface $B = \{ \cup_{t} \Omega_t : t_1 \leq t \leq t_2
\}$.  We then define $M \subset \cM$ as the region ``inside'' $B$ bounded
by the surfaces $\Sigma_{t1}$ and $\Sigma_{t2}$. $\Sigma_t$ and $\Omega_t$
represent foliations of $M$ and $B$ respectively. Figure \ref{M}
illustrates these concepts for a three dimensional $\cM$.

We define unit normal vector fields for the various hypersurfaces. Already
we have defined $u^\az$ as the timelike unit normal vector field to the
$\Sigma_t$ surfaces. Similarly, we may define $\ou^\az$ as the forward
pointing timelike unit normal vector field to the surfaces $\Omega_t$ in
the hypersurface $B$. The spacelike outward-pointing unit normal vector
field to $B$ is defined as $n^\az$.  Then, by construction $\ou^\az n_\az
= 0$ and $T^\az n_\az = 0$. We further define $\on^\az$ as the vector
field defined on $B$ such that $\on^\az$ is the unit normal vector to
$\Omega_t$ in $\Sigma_t$ ($\Omega_t$ viewed as a surface in $\Sigma_t$).
By construction $u^\az \on_\az = 0$.

We define the scalar field $\eta = u^\az n_\az$ over $B$. If $\eta = 0$
everywhere, then the foliation surfaces are orthogonal to the boundary $B$
(the case dealt with in refs. \cite{BY1,HHorowitz} \footnote{The
definitions of $u^\az$ and $n^\az$ are consistent with ref. \cite{BY1}, but
interchanged with respect to those in ref. \cite{HHorowitz}.}), and the
vector fields with the tildes
are equal to their counterparts without tildes. We express $\ou^\az$ and
$n^\az$ in terms of $u^\az$ and $\on^\az$ (or vice versa) as,
\bea
n^\az =\frac{1}{\lambda} \on^\az - \eta u^\az &\mbox{ and }& \ou^\az 
= \frac{1}{\lambda} 
u^\az - \eta \on^\az, \label{nou} \\
&\mbox{or,}& \nn \\
\on^\az =\frac{1}{\lambda} n^\az + \eta \ou^\az &\mbox{ and}& u^\az 
= \frac{1}{\lambda} 
\ou^\az + \eta n^\az, \label{onu}
\eea
where $\lambda^2 \equiv \frac{1}{1 + \eta^2}$. 

Note too, that with $T^\az n_\az = 0$ we may write,
\bea
\label{boundaryTime}
T^\az = \oN \ou^\az + \oV^\az,
\eea
where we call $\oN \equiv \lambda N$ the boundary lapse and $\oV^\az
\equiv \sigma^\az_{\ \bz} V^\bz$ the boundary shift. 

Next consider the metrics induced on the hypersurfaces by the spacetime
metric $g_{\az \bz}$.  These may be written in terms of $g_{\az \bz}$ and
the normal vector fields. $h_{\az \bz} \equiv g_{\az \bz} + u_\az u_\bz$
is the metric induced on the $\Sigma_t$ surfaces, $\gamma_{\az \bz} \equiv
g_{\az \bz} - n_\az n_\bz$ is the metric induced on $B$, and $\sigma_{\az
\bz} \equiv h_{\az \bz} - \on_\az \on_\bz = \gamma_{\az \bz} + \ou_\az
\ou_\bz$ is the metric induced on $\Omega_t$. By raising one index of
these metrics we obtain projection operators into the corresponding
surfaces. These have the expected properties: $h^\az_{\ \bz} u^\bz =
\gamma^\az_{\ \bz}n^\bz = \sigma^\az_{\ \bz} n^\bz = \sigma^\az_{\ \bz}
u^\bz = 0$, and $h^\az_{\ \bz} h^\bz_{\ \cz} = h^\az_{\ \cz}$,
$\gamma^\az_{\ \bz} \gamma^\bz_{\ \cz} = \gamma^\az_{\ \cz}$, and
$\sigma^\az_{\ \bz} \sigma^\bz_{\ \cz} = \sigma^\az_{\ \cz}$. 

On choosing a coordinate system $\{x^1,x^2,x^3\}$ on the surface
$\Sigma_0$ we define $h = \mbox{det}(h_{\az \bz})$ (where in this case we
take $h_{\az \bz}$ as the coordinate representation of that metric tensor).
We then map this coordinate system to each of the other $\Sigma_t$
surfaces using the time flow; combining this set of coordinates on each
surface with the time coordinate $x^0 \equiv t$ we have a coordinate
system over all of $M$. We define $g = \mbox{det} (g_{\az \bz})$.
Similarly, choosing a coordinate system on $\Omega_t$ we define $\sigma =
\mbox{det} (\sigma_{\az \bz})$. Again, using the time flow to extend the
coordinate system over all of $B$, we define $\gamma = \mbox{det}
(\gamma_{\az \bz})$. It is then not hard to show \cite{HHunter} that
\be
\sqrt{-g} = N \sqrt{h} \mbox{ and } \sqrt{-\gamma} = \oN \sqrt{\sigma}.
\label{dets}
\ee

We also define the following extrinsic curvatures.  Taking $\nabla_\az$ as
the covariant derivative on $\cM$ compatible with $g_{\az \bz}$, the
extrinsic curvature of $\Sigma_t$ in $\cM$ is $ K_{\az \bz} \equiv -
h^\cz_{\ \az} h^\dz_{\ \bz} \nabla_\cz u_\dz = - \frac{1}{2} \pounds_u
h_{\az \bz}$, where $\pounds_u$ is the Lie derivative in the direction
$u^\az$. The extrinsic curvature of $B$ in $\cM$ is $\Theta_{\az \bz} = -
\gamma^\cz_{\ \az} \gamma^\dz_{\ \bz} \nabla_\cz n_\dz$, while the
extrinsic curvature of $\Omega_t$ in $\Sigma_t$ is $k_{\az \bz} \equiv -
\sigma^\cz_{\ \az} \sigma^\dz_{\ \bz} \nabla_\cz \on_\dz$. Contracting
each of these with the appropriate metric we define $K \equiv h^{\az \bz}
K_{\az \bz}$, $\Theta \equiv \gamma^{\az \bz} \Theta_{\az \bz}$, and $k
\equiv \sigma^{\az \bz} k_{\az \bz}$.

Finally, we define the following intrinsic quantities over $\cM$ and
$\Sigma_t$. On $\cM$, the Ricci tensor, Ricci scalar, and Einstein tensor
are $\cR_{\az \bz}$, $\cR$, and $G_{\az \bz}$ respectively. On $\Sigma_t$,
$D_\az$ is the covariant derivative compatible with $h_{\az \bz}$, while
$R_{\az \bz}$ and $R$ are respectively the intrinsic Ricci tensor and
scalar.

\section{Analysing the action}

For definiteness, we now take $\cM$ to be four dimensional. The 
generalization to other dimensions is trivial. Given $M \subset \cM$ 
as described above and allowing for the inclusion of a cosmological 
constant, Hayward's action \cite{Hayward} is:
\bea
\label{action}
I &=& \frac{1}{2 \kappa} \int_M d^4 x \sqrt{-g} (\cR - 2 \Lambda) 
+ \frac{1}{\kappa} \int_\Sigma d^3 x 
\sqrt{h} K - \frac{1}{\kappa} \int_B d^3 x \sqrt{- \gamma} \Theta \\ 
&&+ \frac{1}{\kappa} 
\int_\Omega d^2 x \sqrt{\sigma} \sinh^{-1} (\eta) - \bI, \nn 
\eea
where $\int_\Sigma = \int_{\Sigma_2 - \Sigma_1}$ and $\int_\Omega =
\int_{\Omega_2} - \int_{\Omega_1}$, and if we choose a system of units
where
$c = G = 1$, $\kappa = 8 \pi$.  $\bI$ is a functional of the
boundary metrics on $\partial M$. For simplicity, in the next two sections
we will take $\bI = 0$, but in the section \ref{background} we will allow
it to be non-zero again. 

\subsection{Variation of the action}

The variation of $I$ with respect to the metric $g_{\az \bz}$ is \cite{Hayward}
\bea
\label{FirstVar}
\delta I &=& \frac{1}{2 \kappa} \int_M d^4 x \sqrt{-g} (G_{\az \bz} + 
\Lambda g_{\az \bz})  \delta g^{\az \bz} \\
&& - \int_\Sigma d^3 x  \left( P^{\az \bz} \delta h_{\az 
\bz} \right) + \int_B d^3 x \left( \pi^{\az \bz} \delta \gamma_{\az \bz} 
\right) \nn \\
&& +  \int_\Omega d^2 x 
\left( \frac{1}{ \kappa} \sinh^{-1}(\eta) \delta \sqrt{\sigma} \right), \nn
\eea
where $P^{\az \bz} \equiv \frac{\sqrt{h}}{2 \kappa} 
\left( K^{\az \bz} - K h^{\az \bz} \right)$ and 
$\pi^{\az \bz} \equiv \frac{\sqrt{-\gamma}}{2 \kappa} 
\left( \Theta^{\az \bz} - \Theta \gamma^{\az \bz} \right)$. 
If we consider variations of the metric that leave the boundary 
metrics $h_{\az \bz}$ and  $\gamma_{\az \bz}$ fixed, then all of the 
boundary terms are $0$, and $\delta I = 0$ if and only if Einstein's 
equations hold over $M$. 
Thus $I$ is the action that generates general relativity if we are 
considering variations of the metric over a bounded region of space such as $M$.

Now, $\gamma_{\az \bz}$ is fully defined if we specify $\oN$, $\oV^\az$,
and $\sigma_{\az \bz}$. Thus variation of $\gamma_{\az \bz}$ is
equivalent to a variation of these quantities, and we may rewrite the $B$
term in the above with respect to such variations. During this calculation
we repeatedly make use of the fact that $\delta u_\alpha \| u_\alpha$,
$\delta n_\alpha \| n_\alpha$. This is true because these one forms are
defined by the requirement that $u_\az v^\az = n_\az w^\az = 0$ for all
vector fields $v^\az \in T \Sigma_t $, $w^\az \in TB$. The metric does not
figure in the definition of $T \Sigma_t$ or $TB$ nor is a metric required
to calculate the action of a one form on a vector, so $u_\az$ and $n_\az$
are defined up to a normalizing factor independently of the metric.
Thus, $\sigma^{\az \bz} \delta u_\bz = \sigma^{\az \bz} \delta \ou_\bz  = 
\sigma^{\az \bz} \delta n_\bz = \sigma^{\az \bz} \delta \on_\bz = 0$.

Expressing $u^\az$ and $\on^\az$ in terms of $\ou^\az$ and $n^\az$ and writing
$\gamma_{\az \bz} = \sigma_{\az \bz} - \ou_\az \ou_\bz$, we have
\be
\label{p1}
(\Theta^{\az \bz} - \Theta \gamma^{\az \bz})  \delta \gamma_{\az \bz} = 
(\Theta^{\az \bz} - 
\Theta \sigma^{\az \bz}) \delta \sigma_{\az \bz}
+ 2 \Theta_{\az \bz} \ou^\az \delta \ou^\bz
- 2 \Theta \ou_\az  \delta \ou^\az.
\ee
In the meantime, we may decompose $\Theta_{\az \bz}$ into its parts that are
perpendicular and 
parallel to the $\Omega_t$ to obtain
\be
\label{p2}
\Theta_{\az \bz} = \ok_{\az \bz} + 2 \ou_{(\az} 
\sigma_{\bz)}^{\ \cz} \ou^\dz \nabla_\cz 
n_\dz + \ou_\az \ou_\bz  (n^\cz  \oa_\cz),
\ee
where
$\ok_{\az \bz} \equiv - \sigma_\az^{\ \cz} \sigma_\bz^{\ \dz} 
\nabla_\cz n_\dz$ (the extrinsic curvature of $\Omega_t$ in a hypersurface 
perpendicular to $B$) and
$\oa_\az \equiv \ou^\bz \nabla_\bz \ou_\az$ is the acceleration of normal vector 
$\ou^\az$ along its length. Also, on contracting $\Theta^{\az \bz}$ with 
$\gamma_{\az \bz}$,
\be 
\label{p3}
\Theta = \ok - n^\az \oa_\az,
\ee
where $\ok \equiv \sigma^{\az \bz} k_{\az \bz}$.
Putting these three results (\ref{p1},\ref{p2},\ref{p3}) together, 
a few lines of algebra produces
\be
\label{preThetaTerm}
(\Theta^{\az \bz} - \Theta \gamma^{\az \bz} ) \delta \gamma_{\az \bz} =
( \ok^{\az \bz} - [\ok - n^\cz \oa_\cz] \sigma^{\az \bz}) \delta 
\sigma_{\az \bz} + 
2 \sigma_\bz^{\ \cz} n^\dz \nabla_\cz \ou_\dz \delta \ou^\bz - 
2 \ok \ou_\bz \delta
\ou^\bz.
\ee
To complete this deconstruction, recall that the 
time-flow vector field is defined independent of the metric. Therefore $\delta 
T^\az = 0$ which in turn implies that $\delta \ou^\az = -\frac{1}{\oN} 
\left( \ou^\az \delta 
\oN + \delta \oV^\az \right)$. Applying this to equation (\ref{preThetaTerm}), 
substituting the result back into the variation of the action 
(\ref{FirstVar}), and recalling (\ref{dets}) we obtain the final result
\bea
\label{finalVar}
\delta I &=& \frac{1}{2 \kappa} \int_M d^4 x \sqrt{-g} ( G_{\az \bz} + 
\Lambda g_{\az \bz}) \delta g^{\az \bz} \\
 && - \int_\Sigma d^3 x \left( P^{\az \bz} \delta h_{\az 
\bz} \right) + \int_\Omega d^2 x \left( \frac{1}{ \kappa} 
\sinh^{-1}(\eta) \delta 
\sqrt{\sigma} \right) \nn \\
&& - \int dt \int_{\Omega_t} d^2 x \sqrt{\sigma} \left[ \oep \delta \oN - 
\oj_\bz \delta \oV^\bz - \frac{\oN}{2} \os^{\az \bz} \delta \sigma_{\az \bz}
\right] \nn,
\eea
where $\oep \equiv \frac{1}{\kappa} \ok$, $\oj_\az \equiv -\frac{1}{\kappa} 
\sigma_\az^{\ \bz} n^\dz \nabla_\bz \ou_\dz$, and $\os^{\az \bz} 
\equiv \frac{1}{\kappa} ( \ok^{\az \bz} - [\ok - n^\cz \oa_\cz] 
\sigma^{\az \bz})$.

From this result we can make a couple of useful observations. First,
examining the initial and final hypersurfaces $\Sigma_1$ and $\Sigma_2$
and their boundaries $\Omega_1$ and $\Omega_2$ we see that $P^{\az \bz}$
is the $\Sigma_t$ hypersurface momentum conjugate to $h_{\az \bz}$ while
$\frac{1}{2 \kappa} \sinh^{-1} (\eta)$ is the $\Omega_t$ hypersurface
momentum conjugate to $\sqrt{\sigma}$.  Secondly we see that
$-\sqrt{\sigma} \oep$ is conjugate to the boundary lapse $\oN$,
$\sqrt{\sigma} \oj^\az$ is conjugate to the boundary shift $\oV^\az$, and
$\frac{1}{2} \oN \sqrt{\sigma} \os^{\az \bz}$ is conjugate to the boundary
metric $\sigma_{\az \bz}$. Following the Hamilton-Jacobi analysis of
\cite{BY1}, we identify $\oep$, $\oj^\az$, and $\os^{\az \bz}$ as surface
energy, momentum, and stress densities.  If $\eta = 0$, these quantities
coincide with those defined in \cite{BY1}. Also note that each of these
terms is explicitly independent of $\eta$. They are defined with respect to
the foliation of $B$ only. 

\subsection{Decomposing the action}
We now decompose $I$ with respect to the foliation.
To start, the $\cR-2 \Lambda$ term of (\ref{action}) may be rewritten as
\bea
\int_M d^4 x \sqrt{-g} (\cR -2 \Lambda) &=& \int_M d^4 x \sqrt{-g} 
\left( R -2 \Lambda - K^2 + K_{\alpha \beta} K^{\alpha \beta} 
\right) \\
&& - 2 \int_\Sigma d^3 x \sqrt{h} K - 2 \int_B d^3 
x \sqrt{-\gamma} \left( K \eta + n_\az a^\az \right) \nn,
\eea
where $a^\az \equiv u^\bz \nabla_\bz u^\az$ is the acceleration 
of the foliation's unit normal vector field along its length. Next, 
if we substitute expressions for $n^\az$ and $\ou^\az$ from eq. 
(\ref{nou}) into (\ref{p3}) then it is a simple matter to show that,
\be
\Theta + \eta K + n^\az a_\az = \frac{1}{\lambda} k + \lambda \ou^\az 
\nabla_\az \eta.
\ee
Combining these two results the following expression for $I$ is obtained.
\bea
I &=& \frac{1}{2 \kappa} \int_M d^4 x \sqrt{-g} 
\left( R - 2\Lambda - K^2 + K_{\az \bz} K^{\az \bz} \right) \\ 
&& - \frac{1}{\kappa} \int_B d^3 x \sqrt{-\gamma} \left( \frac{k}{\lambda} + 
\lambda \ou^\az \nabla_\az \eta \right) 
+ \frac{1}{\kappa} \int_\Omega d^2 x \sqrt{\sigma} \sinh^{-1} (\eta) \nn.
\eea
Next, we apply the Einstein constraint equations. These are
\bea
\cH &\equiv& - \frac{\sqrt{h}}{\kappa} G_{\alpha \beta} u^\alpha 
u^\beta = -\frac{\sqrt{h}}{\kappa} 
\left( R - 2 \Lambda + K^2 - K_{\az \bz} K^{\az \bz} \right) = 0, \mbox{ and}\\ 
\cH_\az &\equiv& \frac{\sqrt{h}}{\kappa} h_\az^{\ \bz} G_{\bz \cz} u^\cz = 
\frac{\sqrt{h}}{\kappa} \left( D_\bz K_\az^{\ \bz} - D_\az K \right)= 0 .
\eea
Combining these constraints with the Lie derivative definition of the
extrinsic curvature $K_{\az \bz} = - \frac{1}{2} \pounds_{u} h_{\az \bz} =
- \frac{1}{2 N} \left( \pounds_T h_{\az \bz} - 2 D_{(\az} V_{\bz)}
\right)$ of $\Sigma_t$ in $\cM$, we may rewrite the integrand of the
remaining bulk term with respect to these constraints, a time derivative
of the hypersurface metric, and a total divergence term: 
\be
\label{inter1}
R - 2 \Lambda - K^2 + K_{\az \bz} K^{\az \bz} = 
\frac{2 \kappa}{\sqrt{h}} P^{\az \bz} 
\pounds_T h_{\az \bz}  - \frac{2\kappa}{\sqrt{h}} \cH - 
\frac{2 \kappa}{\sqrt{-g}} 
V^\az \cH_\az - \frac{4 \kappa}{N} 
D_\az \left[ \frac{1}{\sqrt{h}} P^{\az \bz} V_\bz 
\right],
\ee
where 
$P^{\az \bz}$ is the hypersurface momentum for $\Sigma_t$ that 
we discussed above.
Then, using Stokes theorem on the hypersurfaces $\Sigma_t$ to move the total 
divergence term out to the boundaries $\Omega_t$ and applying (\ref{dets})
we may write the action as:
\bea
I &&= \int_M d^4 x \left( P^{\az \bz} \pounds_T h_{\az \bz} 
- N \cH - V^\az H_\az \right)\\ 
  && -\frac{1}{\kappa} \int dt \int_{\Omega_t} d^2 x \sqrt{\sigma} 
\left( Nk - V^\az [K_{\az \bz} - K h_{\az \bz} ] \on^\bz - 
\oN \lambda \ou^\az \nabla_\az \eta \right)  \nn \\
  && + \frac{1}{\kappa} \int_{\Omega} d^2 x \sqrt{\sigma} \sinh^{-1} (\eta) \nn.
\eea
Up to this point we have been working with the foliation of $M$ and
therefore with the lapse $N$, shift $V^\az$, and normal vectors $u^\az$
and $\on^\az$. On the term evaluated on $B$ we now switch to work with the
foliation of $B$ and therefore the boundary lapse $\oN$, the boundary shift 
$\oV^\az$, and normal vectors $\ou^\az$ and $n^\az$. Then,
\bea
N k &=& \frac{1}{\lambda^2} \oN \ok - \eta N \sigma^{\az \bz} 
\nabla_\az \ou_\bz \\
 -\on^\az \left( K_{\az \bz} - K h_{\az \bz} \right) V^\bz &=& N \eta
 \sigma^{\az \bz} \nabla_\az \ou_\bz - \oN \eta^2 \ok + n^\az \oV^\bz 
\nabla_\bz \ou_\az + \lambda \oV^\bz \nabla_\bz \eta,
\eea
where
$\ok \equiv \sigma^{\az \bz} \ok_{\az \bz}$.
Combining these two results with (\ref{boundaryTime}), which can be used 
to show that
\bea
&& \int_\Omega d^2 x \sqrt{\sigma} \sinh^{-1}(\eta) - \int dt 
\int_{\Omega_t} d^2 x \sqrt{\sigma} \lambda \oN \ou^\az  \nabla_\az \eta \\
&=& \int dt \int_{\Omega_t} d^2 x \left( (\pounds_T \sqrt{\sigma}) 
\sinh^{-1}(\eta) + \sqrt{\sigma} \lambda \oV^\az \nabla_\az \eta \right) \nn,
\eea
we obtain the following decomposition of the action,
\bea
\label{decomp}
I &=& \int_M d^4 x \left( P^{\az \bz} \pounds_T h_{\az \bz} 
- N \cH - V^\az H_\az \right)
      + \frac{1}{ \kappa} \int dt \int_{\Omega_t} d^2 x 
(\pounds_T \sqrt{\sigma}) \sinh^{-1}(\eta) \\
  && - \int dt \int_{\Omega_t} d^2 x \sqrt{\sigma} 
\left( \oN \oep - \oV^\az \oj_\az \right) 
\nn, 
\eea  
where $\oep$ and $\oj^\az$ are the energy and surface momentum densities
that we obtained from the variational calculation. 

The terms of this expression will be familiar to anyone who is familiar
with refs. \cite{BY1,HHorowitz,HHunter}. Specifically, $\oep$ and
$\oj^{\alpha}$ are exactly the energy surface density and momentum surface
density that the observers on the boundary would measure if the foliation
of $M$ were perpendicular to $B$. A little thought shows that these
quantities are the ones that would be reasonable for observers restricted
to surface $B$ to measure. Such observers are cognizant of the foliation
of the boundary (for the foliation has been defined to correspond to their
notion of simultaneity), but being restricted to the surface they have no
way of associating that foliation with a foliation of $M$ as a whole.
Viewed another way, there are no observers in the interior of $M$ and
therefore no unique way to extend the ``instants'' of time into that
region. As such, it does not seem to physically make sense for the
observers to measure the energy and momentum surface densities with
respect to the foliation $\Sigma_t$, the lapse $N$, and the shift $V^\az$
that we have defined but they cannot observe. Rather, as observers
travelling along $B$ they would naturally (locally) extend the foliation
of $B$ into a foliation that is perpendicular to $T^\az$ - and therefore
in effect be considering a foliation $\tilde{\Sigma_t}$ (locally) defined
around $B$ with lapse $\oN$ and shift $\oV^\az$. Then they would measure
the quantities that we have found naturally arise from the action.

We may also define a Hamiltonian. In elementary classical mechanics with
one degree of freedom, the action $I$ and Hamiltonian $H$ are related by
the equation $I = p \dot{q} - H$, where $q$ is the variable giving the
configuration of the system and $p = \frac{\partial I}{\partial q}$ is the 
conjugate momentum. Extending this to the system under consideration 
\cite{BY1}, $h_{\az \bz}$ and $\sqrt{\sigma}$ are configuration variables 
while $P^{\az \bz}$ and $\frac{1}{ \kappa} \sinh^{-1}(\eta)$ are their 
conjugate momenta and so the Hamiltonian is:
\bea
H = \int_{\Sigma_t} d^3 x [N \cH + V^{\az} H_{\az}] + \int_{\Omega_t} 
d^2 x \sqrt{\sigma} (\oN \oep - \oV^\az \oj_{\az}).
\eea
Again this quantity is indifferent to the intersection angle between the
foliation of $M$ and the boundary. For solutions to the Einstein
equations, it is defined entirely with respect to the foliation of $B$.
Note that this Hamiltonian does not agree with that proposed in
\cite{HHunter} where the problem was approached from the point of view of
the foliation of $M$ rather than that of $B$. 

\subsection{Conserved Charges}

The discussion of conserved charges presented in \cite{BY1} carries over
exactly into this work. Thus if $\xi^\az \in TB$ is a vector field in the
boundary $B$ and $\pounds_\xi \gamma_{\az \bz} = 0$ (ie. it is a Killing
vector field), then we may define an associated conserved charge
\be
Q_\xi \equiv \int_\Omega d^2 x \sqrt{\sigma} \xi^\az 
\left( \oep \ou_\az + \oj_\az \right).
\ee 
If $T^\az$ is a Killing vector field then the 
Hamiltonian $H$ as defined above is a conserved charge. If there is an 
angular Killing vector field $\phi^\az \in T\Omega$ then the angular momentum
\be
J_\phi \equiv \int_\Omega d^2 x \sqrt{\sigma} \phi^\az \oj_\az,
\ee
is also a conserved charge. 

\subsection{Background Terms}
\label{background}

We now return to the reference term $\bI$. Defined as it is as a
functional of the boundary metrics, it is clear that for a metric
variation that leaves the boundary metrics unchanged, $\delta \bI=0$ -
therefore its exact form does not affect the equations of motion.  This
degree of freedom in the definition of $I$ may equivalently be viewed as
the freedom to define zero points of the energy, momentum, and
Hamiltonian. Specifically, it allows us to choose a reference spacetime
for which we wish these quantities to be zero. For asymptotically flat
spacetimes we would normally choose Minkowski space as the reference
spacetime, but other choices may be made if we are studying spacetimes
with other asymptotic behaviours - for example asymptotically
anti-deSitter space \cite{BCM}. 

Given a reference spacetime $(\underline{M},\underline{g}_{\az \bz})$, we
embed $(\Omega,\sigma_{\az \bz})$ in that spacetime and define a vector
field $\bT^\az$ over the embedded
$(\underline{\Omega},\underline{\sigma}_{\az \bz})$ such that $\bT^\az \bT_\az =
T^\az T_\az$ and the components of $\pounds_{\bT} \underline{\sigma}_{\az
\bz} = \pounds_T \sigma_{\az \bz}$ \footnote{We leave aside the issue as
to whether this is possible in all cases. We will consider several
examples where it is but in general an arbitrary surface cannot be
embedded in an arbitrary higher dimensional space.}. These conditions
ensure that the boundary lapse and the components of the boundary shift
vector as calculated
from $\bT^\az$ are equal to those calculated for $T^\az$ in the original
spacetime. We then define
\be 
\bI = \int dt \int_{\Omega} d^2 x \sqrt{\sigma} 
[\oN \underline{\oep} - \oV^\az \underline{\oj}_\az ],
\ee
where $\underline{\oep}$ and $\underline{\oj}_\az$ are defined in the same
way as before except that this time they are evaluated for the surface
$\Omega$ embedded in the reference spacetime.  Thus, the net effect of
including $\bI$ is to change $\oep \rightarrow \oep - \underline{\oep}$
and $\oj_\az \rightarrow \oj_\az - \underline{\oj}_\az$. 

Physically these conditions correspond to demanding that an observer
living in the surface $\Omega$ and observing only quantities intrinsic to
that surface (as it evolves through time) cannot tell whether she is
living in the original spacetime or in the reference spacetime. 
From another point of view the observers have calibrated their instruments
so that they will always measure the quasilocal quantities to be zero in
the reference spacetime - no matter what kind of motion they undergo. 
 
This definition of $\bI$ differs slightly from both the one used in
\cite{BY1} and the one used in \cite{HHunter}. In the former case $\Omega$
was embedded in a reference three dimensional space and no demand was made
of $\pounds_T \sigma_{\az \bz}$; however in all examples considered in
ref. \cite{BY1} (and indeed in the subsequent work) $\pounds_T \sigma_{\az
\bz}=0$, so insofar as that formalism has been pursued within the
literature it agrees with the formalism considered here.  Note that if we
do not include the conditions on $\pounds_T \sigma_{\az \bz}$ then boosted
observers in Minkowski space will observer non-zero quasilocal energies
which is clearly an undesirable situation.

In ref. \cite{HHunter} $(B,\gamma_{\az \bz})$ as a whole is embedded in
the reference four space $(\underline{M},\underline{g}_{\az \bz})$. That
requirement is essentially the global version of our definition of $\bI$
and as such will locally yield the same results as our definition though
it is somewhat harder to apply computationally. Beyond that condition they
further require that the reference spacetime be foliated in such a way
that $\eta$ in the reference spacetime is the same as in the original
spacetime. Such a condition is neither necessary nor desirable
in our approach which doesn't concern itself with the foliation of the
spacetime as a whole. Finally we note that in the approach used in ref.
$\cite{HHunter}$ the inclusion of this background term is necessary to
remove an $\eta$ dependence in the Hamiltonian - this dependence does not
occur in our approach.

\section{Examples}

We now consider some sample calculations. For simplicity we will work with
static spherically symmetric spacetimes parameterized with the natural spherical
coordinates $\{t,r,\theta,\phi \}$ and therefore with metric
\be
ds^2 = -F(r) dt^2 + \frac{ dr^2 }{F(r)} 
+ r^2 ( d\theta^2 + \sin^2 \theta d \phi^2 ).
\ee
In each case we will consider a surface of observers $\Omega$ 
defined by $r=r_0$ and $t=t_0$. Then geometrically $\Omega$ it is a 
two sphere with metric
\be
ds^2 = r_0^2 (d \theta^2 + \sin^2 \theta d \phi^2 ),
\ee
(where we have parameterized $\Omega$ with the same $\theta$ and $\phi$ 
as the full space).
If we then consider the timelike unit vector field,
\be
T^\az = \left[ \frac{\oN}{\sqrt{F}} \sqrt{ 1 + \frac{R^2}{\oN^2 F} }, 
R, \Theta, \Phi \right],
\ee
where $R=R(r,\theta,\phi)$, $\Theta=\Theta(r,\theta,\phi)$, and 
$\Phi = \Phi(r,\theta,\phi)$ are general functions of 
$r,\theta, \mbox{ and } \phi$, the boundary lapse and boundary shift 
functions are easily found to be
\be
\oN^2 = 1 + r^2 (\Theta^2 + \sin^2 \theta \Phi^2 ),
\ee
and
\be
\oV^\az = [0,0,\Theta,\Phi],
\ee
while
\be
\ou^\az = \left[ \frac{1}{\sqrt{F}} \sqrt{ 1 + \frac{R^2}{\oN^2 F} }, 
\frac{R}{\oN}, 0 , 0 \right],
\ee
and
\be
n_\az = \left[ - \frac{R}{\oN}, \frac{1}{\sqrt{F}} 
\sqrt{ 1 + \frac{R^2}{\oN^2 F} }, 0, 0 \right].
\ee
Then, a straightforward calculation yields
\bea
\oep &=& - \frac{2}{\kappa r} \sqrt{ F + \frac{R^2}{\oN^2} }, \mbox{ and}\\
\oj_\az &=& -\frac{2}{\kappa \sqrt{F + \frac{R^2}{\oN^2} } } 
\left[ 0,0, \frac{\partial}{\partial \theta} \left( \frac{R}{\oN} \right), 
\frac{\partial}{\partial \phi} \left( \frac{R}{\oN} \right) \right].
\eea
Finally, we calculate the $\Omega$ components of 
$\pounds_T \sigma_{\az \bz}$ as  
\bea
\left( \pounds_T \sigma_{\az \bz} \right)_{\theta \theta} 
& = & 2 r \left( R + r \frac{\partial \Theta}{\partial \theta} \right), \\
\left( \pounds_T \sigma_{\az \bz} \right)_{\theta \phi} 
& = & r^2 \left( \frac{\partial \Theta}{\partial \phi} 
+ \sin^2 \theta \frac{\partial \Phi}{\partial \theta} \right), \\
\left( \pounds_T \sigma_{\az \bz} \right)_{\phi \phi} 
& = & 2 r \sin \theta \left( R \sin \theta  + \Theta r \cos \theta  
+  r \sin \theta \frac{\partial \Phi}{\partial \phi} \right).
\eea
Note that there is no $F$ dependence in any of these components. Thus if
we wish to calculate quasilocal quantities for observers moving through
Schwarzschild space using Minkowksi space as a reference spacetime (as in
the following examples), on embedding $\Omega$ 
(which is trivial for a sphere) and setting
\be 
\bT^\az = \left[ \oN \sqrt{ 1 + \frac{R^2}{\oN^2} } , R, \Theta, \Phi \right],
\ee
the metric $\sigma_{\az \bz}$ and its derivative $\pounds_T \sigma_{\az
\bz}$ will be the same for both Schwarzschild and Minkowski space. 

We now specialize to specific examples using the system of units where
$\kappa = 8 \pi$. 

\subsection{Static Observers} 
For our first example, we'll consider a
spherical set observers holding themselves static with respect to a
Schwarzschild black hole ($F= 1 - \frac{2m}{r}$) and take flat Minkowski
space ($F=1$) as our reference spacetime. Then for both spacetimes
$R=\Theta=\Phi = 0$, $\oN = 1$, and $\oV^\az = 0$. Substituting these data
into the above expressions we obtain
\be
\oep - \underline{\oep} = \frac{1}{4 \pi r} 
\left(1 -  \sqrt{1 - \frac{2m}{r} } \right). 
\ee
Then, the total measured energy is
\be
E = \int_\Omega d^2 x \sqrt{\sigma} (\oep - \underline{\oep}) = 
r \left(1 - \sqrt{ 1 - \frac{2m}{r} } \right).
\ee
This is the standard result as obtained in \cite{BY1}. 
Taking the limit as $r \rightarrow \infty$ we obtain 
$E \rightarrow m$ as would be expected, while as 
$r \rightarrow 2m$ (the Schwarzschild horizon), $ E \rightarrow 2m$. 

With $\oN=1$ and the shift vector $0$, the Hamiltonian $H=E$. 
Since $T^\az$ is a Killing vector in this case, $H$ is a conserved charge. 
There is also a conserved angular momentum associated with each of the regular 
three spherical Killing vectors. $\oj_\az$ is zero however, so each of 
these charges vanishes.

\subsection{Radially Infalling Observers}

A more interesting example is the case of observers taking their
measurements as they fall radially along geodesics towards a Schwarzschild
hole. Such motion is described by solutions to the geodesic equation. For
observers who were stationary as they started falling in from infinity, 
the geodesic equation reduces to
$\frac{dr}{d \tau} = -\sqrt{\frac{2m}{r}}$, where $\tau$ is the proper 
time coordinate. Then,
$R =  -\sqrt{\frac{2m}{r}}$, $\Theta=\Phi = 0$, $\oN = 1$, 
again $\oV^\az = 0$, and
\be
\oep - \underline{\oep} = \frac{1}{4 \pi r} \left( \sqrt{ 1 + \frac{2m}{r} } 
- 1 \right).
\ee
The total measured energy is,
\bea
E = \int_\Omega d^2 x \sqrt{\sigma} (\oep - \underline{\oep}) = 
r \left( \sqrt{ 1 + \frac{2m}{r} } - 1 \right).
\eea
As for static observers as $r \rightarrow \infty$, $E \rightarrow m$. 
Of course, this is not really surprising since the radially infalling 
observers at infinity actually are static! 
Over the rest of the range the two energy measures are not the same. 
In particular as $r \rightarrow 2m$, $E \rightarrow 2m(\sqrt{2} - 1)$. 

As for the first example, the momentum terms are zero and the $\oN=1$, and
so $H=E$. $T^\az$ is no longer a Killing vector however, so this is no
longer a conserved charge. Physically of course this is to be expected
since the observers are moving radially inwards and therefore through the
gravitational field. As time passes therefore the amount of gravitational
field energy contained within $\Omega_t$ changes.  Again the three angular
momenta are conserved but each has the uninteresting value of zero. 

\subsection{Radially Boosted Observers}

We next consider a set of observers who are boosted to travel radially
with ``constant'' velocity $v$. By constant velocity here we mean that a
second set of observers dwelling on a $t=\mbox{constant}$ surface and
being evolved by the timelike vector field $[1,0,0,0]$ will measure the
first set as having velocity $v$ and acceleration $0$. 

Then $R = \gamma v \sqrt{1 - \frac{2m}{r} }$ ($\gamma =
\frac{1}{\sqrt{1-v^2}}$ - the standard Lorentz factor from special
relativity), $\Theta=\Phi=0$, $\oN=1$, and once more $\oV^\az = 0$. A
simple calculation then obtains,
\be
\oep - \underline{\oep} = \frac{\gamma}{4 \pi r} 
\left( \sqrt{ 1 - \frac{2mv^2}{r}} - \sqrt{1 - \frac{2m}{r}}   \right), 
\ee
and the total measured energy is,
\bea
E = \int_\Omega d^2 x \sqrt{\sigma} (\oep - \underline{\oep}) = 
\gamma r \left( \sqrt{ 1 - \frac{2mv^2}{r}} - \sqrt{1 - \frac{2m}{r} }  \right).
\eea
In this case as $r \rightarrow \infty$, $E \rightarrow \sqrt{1-v^2} m$, 
while as $r \rightarrow 2m$, $E \rightarrow 2m$.

Once more with $\oN = 1$ and $\oV^\az = 0$, $H=E$. Again $T^\az$ is not a 
Killing vector, so this is not a conserved charge. The angular momenta are 
conserved charges though again each is zero.

At first glance it may seem unusual that in the $r \rightarrow \infty$ limit 
$E \propto \frac{m}{\gamma}$. Extrapolating from special relativity we would 
perhaps expect $E \propto \gamma m$. Physically however, it is clear that 
there is a flow of gravitational field energy through the surface $\Omega$. 
That is, there is a $j_\vdash$ component of the momentum. This momentum 
may be seen as ``drawing off'' some of the energy. We will not investigate 
the issue further in this paper, though it is addressed in 
the last example of \cite{HHunter} to a certain extent by the invariant 
quantities defined in \cite{Lau}.

\subsection{Z-Boosted Observers}

Finally we consider a set of observers who are boosted to travel 
``in the $z$-direction'' with ``constant'' velocity $v$. 
By constant we again mean with respect to other
observers who are dwelling on $t=\mbox{constant}$ surfaces and 
being evolved by the timelike vector field $[1,0,0,0]$. Then, 
$R=\gamma v \cos \theta \sqrt{1-\frac{2m}{r}}$, 
$\Theta = \frac{\gamma v \sin \theta}{r}$, $\Phi=0$, $\oN = \sqrt{1+ \gamma^2 v^2 
\sin^2 \theta}$, and $V^\az = [0,0,\frac{\gamma v \sin \theta}{r},0]$. 
We now have
\be
\oep - \underline{\oep} = \frac{1}{4 \pi r}  
\frac{1}{\sqrt{1 - v^2 \cos^2 \theta}}  
\left( \sqrt{ 1 - \frac{2mv^2 \cos^2 \theta}{r}} 
- \sqrt{1 - \frac{2m}{r}} \right), 
\ee
and while $\oj_\phi = \underline{\oj}_\phi= 0$, 
\be
\oj_\theta - \underline{\oj}_\theta = 
\frac{ v \sin \theta}{8 \pi (1-v^2 \cos^2 \theta)}
\left( \sqrt{ \frac{1-\frac{2m}{r}}{1 - \frac{2mv^2 \cos^2 \theta}{r} } }
- 1 \right). 
\ee
In this case, the $\oep - \underline{\oep}$ doesn't integrate into a nice 
tidy form as it did in previous examples. Instead we will consider 
the two usual limiting cases. 
For $r \rightarrow \infty$,
\be
\left( \oep - \underline{\oep} \right)_{r \rightarrow \infty}  
= \frac{m}{4 \pi r^2} \sqrt{1 - v^2\cos^2 \theta  }
\mbox{ and }
\left( \oj_\az - \underline{\oj}_\az \right)_{r \rightarrow \infty} 
= \frac{mv \sin \theta}{4 \pi   }.
\ee
Then, integrating over the $\{t=0, r \rightarrow \infty \}$ two 
surface we obtain,
\bea
E_\infty &=& \int_\Omega d^2 x \sqrt{\sigma} (\oep - \underline{\oep}) \nn \\
&& = \frac{m}{2} \left( \sqrt{1-v^2} + \frac{\arcsin v}{v} \right), 
\mbox{ and } \\
H_\infty &=& \int_\Omega d^2 x \sqrt{\sigma} 
\left( \oN (\oep - \underline{\oep})
- \oV^\az (\oj_\az - \underline{\oj}_\az ) \right) \nn \\ 
&& = \sqrt{1-v^2} m.
\eea
Meanwhile for $r \rightarrow 2m$,
\be
\left( \oep - \underline{\oep} \right)_{r \rightarrow 2m}  
= \frac{1}{4 \pi (2m)} \mbox{ and }
\left( \oj_\az - \underline{\oj}_\az \right)_{r \rightarrow 2m} 
= \frac{v \sin \theta }{4 \pi (1-v^2 \cos^2 \theta) }.
\ee
Integrating over the $\{ t=0,r=2m \}$ two surface we obtain,
\bea
E_{2m} &=& \int_\Omega d^2 x \sqrt{\sigma} (\oep - \underline{\oep}) \nn \\
&& = 2m , \mbox{ and } \\
H_{2m} &=& \int_\Omega d^2 x \sqrt{\sigma} \left( \oN (\oep - \underline{\oep})
- \oV^\az (\oj_\az - \underline{\oj}_\az ) \right) \nn \\ 
&& = m \left( 1 - 2 \gamma 
+ \frac{\gamma}{v} \arcsin v + \frac{2}{\gamma v} \mbox{arctanh }  v \right).
\eea

As in the previous cases $T^\az$ is not a Killing vector and so $H$ is not
a conserved charge, as we would physically expect. Note that in this
situation we have a non-zero component of $\oj_\az$. Despite this we still
do not have any non-zero conserved angular momenta. To see this recall
that the three linearly independent spherical Killing
vectors
are $\phi_1^\az = [0,0,0,1]$, $\phi_2^\az =
[ 0,0,\sin \phi, \cos \phi \cot \theta]$,
and $\phi_3^\az = [0,0,\cos \phi, - \sin \phi \cot \theta]$. Then
$\phi_1^\az \oj_\az =0$, $\phi_2^\az j_\az$ is
proportional to $\sin \phi$, and $\phi_3^\az$ is proportional to $\cos
\phi$. Therefore, as would be expected physically and by symmetry, on
integration $J_{\phi_1} = J_{\phi_2} = J_{\phi_3} = 0$. 

The comments made in the previous example regarding the fact that
$H_\infty = \frac{m}{\gamma}$ apply here as well. 

\section{Discussion}

In this paper we have seen that the orthogonality restriction of \cite{BY1}
may be lifted without too much difficulty. Further, by concentrating on
the foliation of the boundary $B$ rather than the spacetime region $M$ we
avoid many of the technical complications of the non-orthogonal treatment
of \cite{HHunter}, and obtain definitions of quasilocal quantities that
are manifestly independent of the intersection angle between the foliation
of $M$ and the boundary $B$ independent of our choice of the background 
spacetime.

In our choice of how to calculate the reference term $\underline{I}$ on
the
background spacetime we have given local conditions that modify those of
\cite{BY1} in a way that is more appropriate if we are considering moving
observers. These conditions at the same time
remain simpler and easier to implement than those required in \cite{HHunter}. 

In general the Hamiltonian and quasilocal quantities such as angular
momenta are dependent on the motion of the observers as we have seen in
several examples. One somewhat counterintuitive observation is that the
observed mass of a source decreases rather than increases with the motion
of the observers who are measuring that mass. This is a consequence of
choosing our observers in such a way that there is a net flow of
gravitational field energy through the surface $\Omega$.

\section{Acknowledgments} 

This work was supported by the National Science and Engineering Research
Council of Canada. The calculations for the examples were done with the
help of the GRTensorII package \cite{GRT} for Maple.

\end{document}